\documentclass[12pt]{article}    
\usepackage[ansinew]{inputenc} 
\usepackage{hyperref}
\usepackage{graphics}
\usepackage{epsfig}
\usepackage{epstopdf}
\usepackage{graphicx} 
\usepackage{amssymb}  
\usepackage{amsmath}
\usepackage{multirow} 
\usepackage{enumitem} 
\usepackage{color}     
\usepackage{indentfirst}  
\usepackage{tabularx,booktabs}
\usepackage{caption} 
\usepackage{appendix}
\renewcommand\appendix{\section*{Appendix}}
\usepackage[gen]{eurosym}
\usepackage[scientific-notation=true]{siunitx}
\usepackage{epstopdf}

\usepackage[all]{xy}
\usepackage{bm}

\usepackage{lscape}
\usepackage{afterpage}

\begin{document}
\newtheorem{definition}{Definition}
\newtheorem{theorem}{Theorem}
\newtheorem{example}{Example}
\newtheorem{corollary}{Corollary}
\newtheorem{lemma}{Lemma}
\newtheorem{proposition}{Proposition}
\newtheorem{remark}{Remark}
\newenvironment{proof}{{\bf Proof:\ \ }}{\qed}
\newcommand{\qed}{\rule{0.5em}{1.5ex}}
\newcommand{\bfg}[1]{\mbox{\boldmath $#1$\unboldmath}}

\begin{center}

\section*{The risk of death in newborn businesses during the first years in market}

\vskip 0.2in {\sc \bf Faustino Prieto$^a$
\footnote{Corresponding author.\\
E-mail addresses: faustino.prieto@unican.es (F. Prieto),
sarabiaj@unican.es (J.M. Sarabia),
enrique.calderin@unimelb.edu.au (E. Calder\'{\i}n-Ojeda).}, 
Jos\'e Mar\'{\i}a Sarabia$^b$, Enrique Calder\'{\i}n-Ojeda$^c$
\vskip 0.2in

{\small\it
$^a$Department of Economics, University of Cantabria, Santander, Spain\\
$^b$Department of Quantitative Methods, CUNEF University, Madrid, Spain\\
$^c$Centre for Actuarial Studies, Department of Economics, University of Melbourne, Melbourne, Australia
}\\
}

\end{center}

\begin{abstract}\noindent
In this paper, we analyzed how business age and mortality are related during the first years of life, and tested the different hypotheses proposed in the literature.
For that, we used data on U.S. business establishments, with 1-year resolution in the range of age of 0-5 years, in the period 1977-2016, published by the United States Census Bureau.
First, we explored the adaptation of classical techniques of survival analysis (the Life Table and Peto-Turnbull methods) to the business survival analysis. Then, we considered nine parametric probabilistic models, most of them well-known in reliability analysis and in the actuarial literature, with different shapes of the hazard function,
that we fitted by maximum likelihood method and compared with the Akaike information criterion. Our findings show that newborn firms seem to have a decreasing failure rate with the age during the first five years in market, with the exception of the first months of some years in which the risk can rise.
\end{abstract}

\noindent {\bf Key Words}: Newborn firm survival; Failure rate;
Business dynamism.

\section{Introduction}
\label{intro}
During the first years of life, simply surviving is one of the key challenges for new small businesses.
New ventures can fail for many reasons \cite{Everett1998,Boyer2014,Lukason2015,Kato2015,Che2017,Cabrer2018,Zhang2018}, both internal and external (poor management decisions, inadequate financing, insufficient demand, regulations, etc), and have an early death.
An open question is how business age and mortality are related during those first years, in other words, how the risk of death in newborn businesses changes with the age during those crucial first years in market.

Different hypotheses, about the relationship between failure risk (risk of death) and firm age, has been suggested previously in the literature \cite{Bruderl1990}. Among the most known are: {\it The Liability of Newness, of Adolescence, of Senescence, The Liability of Obsolescence} and the age-independent risk hypotheses.
The first one, The Liability of Newness \cite{Stinchcombe1965,Freeman1983}, suggests that the risk of business failure declines with the firm age. Under that hypothesis, aging confers an advantage for living longer due to a decreasing influence with time of some negative factors on business.
The second one, The Liability of Adolescence \cite{Fichman1991,Bhattacharya2015}, suggests that the risk of business failure rises at the beginning as initial resources are consumed, until that risk reaches a maximum and declines afterwards with the firm age.
The next two hypotheses, The Liability of Senescence \cite{Henderson1990,Barron1994} and The Liability of Obsolescence \cite{Sorensen2000}, both suggest that the risk of business failure rises with the firm age, respectively due to agility loss and to the increasing divergence between core competencies and market demands, that make firms less able to adapt to change. Finally, the age-independent risk hypothesis \cite{Daepp2015} suggests that the risk of business failure is approximately constant, approximately independent of the business age. Therefore, considering all those hypotheses, what to expect as an entrepreneur during the first years in business? How will aging affect the chance of survival of a new small business during those first years of life?

In relation to those previous hypotheses, a relevant question is how the time a business was born affects its chances of success.  In that context, an analysis of risk of death of businesses, based on the distribution of firm lifespans over a long period of years all together, would group businesses with different failure pattern, (the longer period chosen, the more heterogeneous grouping), and the results obtained would indicate us only an average behavior. For that, in this paper, we conduct the analysis on a yearly basis, by grouping businesses by their year of birth.

A wide employed technique to measure the risk of business failure is the hazard function
\cite{Kiefer1988,Garrod1990,Holmes2010}.
That technique is based on the actuarial concept of  {\it force of mortality}, that could be defined, in this context, as the instantaneous probability of death of a business at time $t$ (in the interval $t$ to $t+dt$) given that the business has survived up until that time $t\;$ \cite{Makeham1867,Davis1952,Steffensen2016}. That concept can be represented by $h(t)dt=f(t)dt/S(t)$, where $h(t)$ is that hazard function (often called as hazard rate, failure rate or conditional failure rate), and $S(t), f(t)$ are the survival and probability density functions, respectively. The shape of the hazard function $h(t)$ can help us to test, empirically, the diverse hypotheses proposed to explain the relationship between failure risk and business age: the first one, the Liability of Newness hypothesis, would correspond with a monotonically decreasing hazard function, in other words, a decreasing failure rate that can be denoted as {\it DFR}; the second one, the Liability of Adolescence hypothesis, would correspond with an inverted U-shaped hazard function, an upside-down bathtub-shaped failure rate, denoted as {\it Unimodal} or as {\it UBT}; the next two hypotheses, the Liability of Senescence and Obsolescence hypotheses, would correspond with a monotonically increasing hazard function, that means an increasing failure rate, denoted as {\it IFR}; and the last hypothesis considered, the hypothesis of risk of business failure age-independent would correspond with a constant hazard function, a constant failure rate that can be denoted as {\it Constant} or as {\it CFR}.

This study aims to provide empirical evidence of how newborn small businesses die during their first five years of life, in particular, of how their mortality (failure) rates are during that initial period of time.  For this purpose,
we used data of US business, with 1-year resolution in the range of age of 0-5 years, in the period 1977-2016, published by the United States Census Bureau. In order to carry out our analysis, we use two different methods for obtaining the non-parametric estimates for the survival function: a modified Life Table method adapted to the peculiarities of the business survival analysis, and an analytical Peto-Turnbull method with closed-form thanks
to our interval-censored data sample is given through a set of non-overlapping intervals. Then, we considered nine parametric probabilistic models, most of them well-known in reliability analysis and in the actuarial literature, with different shapes of the hazard function (DFR, UBT, IFR, CFR, etc.), that we estimated by maximum likelihood method and compared by the Akaike's information criterion. Finally, for the models that yield the better measures goodness of fit, the shape of the models is analyzed. 

The rest of this paper is organized as follows: in Section \ref{sec:2}, we describe the dataset used; in Section \ref{sec:3}, we give full details of the methods chosen for this research; in Section \ref{sec:4}, we provide empirical evidence of a decreasing failure rate with the firm age during those first five years, with except for the first months of some years in which the risk can rise; finally, conclusions are given in Section \ref{sec:5}.

\section{The data}
\label{sec:2}

In this paper, we used the US Census Bureau's Business Dynamics Statistics (BDS) database, a public-use dataset published on the United States Census Bureau website \cite{BDS2019}. In particular, we used the Establishment Age tables (Establishment Characteristics Data Tables), available for the period 1977-2016, within their 2014, 2015 \& 2016 BDS Releases.

Those tables provide, at the establishment level, the age distribution of US establishments for each year of that period, with 1-year resolution in the range of age of 0-5 years and with 5-year resolution in the range of age of 6-25 years. For each $Year$,  on March 12th, the column $Estabs$ gives the number of active (with positive employment) establishments for each value of $Estab\_Age$ (age of the establishments). In addition, the columns $Estabs\_Entry$ and $Estabs\_Exit$ give, respectively, the number of establishments born or died during the last 12 months. It can be noted that $Estabs\_Entry$ often has positive values for $Estab\_Age>0$ in the dataset, that can be interpreted as corresponding to active establishments not reported in previous years.

For this study, we selected the US establishments with ages between 0 and 5 years (with 1-year time resolution), and we considered 35 different birth cohorts, from 1977 to 2011, that we analyzed separately during their first 5 years of life.
To do this, we followed the notation in \cite[section 3.6.1]{Lawless2011}, and we denoted each time interval as $I_j=[a_{j-1},a_j),\;j=1,\dots,k+1$, with $k+1=6$ intervals \{[0,1),[1,2),\dots,[5,$\infty$)\} in years. Also, we denoted the number of active business at the start of the interval $I_j$ as $N_j$ (given by $Estabs$ in our dataset) and the number of deaths during that interval $I_j$ as $D_j$ (given by $Estabs\_Exit$). In addition, we denoted the number of new entrants during each interval $I_j$ as $E_j$ (given by $Estabs\_Entry$ in our dataset).

As an example, Table \ref{tab:1} shows the last cohort available in our dataset, the 1-year birth cohort 2011 (on the right), and the data in its BDS original format from \cite{BDS2019} corresponding to the period 2011-2016 (on the left). First, there were 522626 newborn establishments alive on March 12, 2011 (see that $Estabs=522626$ for $Estab\_Age=0$ and $Year=2011$), thus, at the start of the first ($j=1$) on-March-to-March interval $I_1=[0,1)$ of the 2011 birth cohort, there were $N_1=522626$ active establishments. Then, from March 12th 2011 to March 12th 2012, there were  $Estabs\_Entry=0$ new entrants and $Estabs\_Exit=100616$ deaths
(see for $Estab\_Age=1$ and $Year=2012$), consequently, the number of new entrants and the number of deaths during the first 1-year interval $I_1=[0,1)$ of the 2011 birth cohort were $E_1=0$ and $D_1=100616$, respectively. Next, there were $Estabs=416101$ establishments alive on March 12, 2012 (see for $Estab\_Age=1$ and $Year=2012$), and thus, $N_2=416101$, and so on.
Finally, there were $Estabs=277495$ establishments alive on March 12, 2016, and thus, corresponding to the interval $I_6=[5,\infty)$ (in years), we have that $N_6=D_6=277495$.

\begin{table}[htp]\footnotesize
\renewcommand{\tablename}{\footnotesize{Table}}
\setlength{\tabcolsep}{4.7pt}
\caption{\label{tab:1}\footnotesize Observed data in its BDS original format (on the left), and the corresponding 2011 birth cohort (on the right): number of active establishments ($N_j$) at the start of each 1-year interval  ($I_j$), and number of new entrants ($E_j$) and deaths ($D_j$) in that interval. Source: US Census Bureau 2014, 2015 \& 2016 BDS Releases.}
\begin{tabular}{c c c c c | c c c c c c}
\hline\noalign{\smallskip}
\multicolumn{5}{c}{Observed Data (Dataset)}&&
\multicolumn{5}{c}{1-Year Birth Cohort 2011}\\
\noalign{\smallskip}\hline\noalign{\smallskip}
\multirow{2}{*}{$Year$} &  $Estabs$ &  \multirow{2}{*}{$Estabs$} & $Estab$ & $Estabs$ &
& \multirow{2}{*}{$j$} &  $I_j${\tiny(in years)} &
\multirow{2}{*}{$E_j$} & \multirow{2}{*}{$D_j$} & \multirow{2}{*}{$N_j$}\\
& $Age\;${\tiny(years)}    &       &  $Entry$    &  $Exit$&      &  &  $[a_{j-1},a_j)$ & & & \\
\noalign{\smallskip}\hline\noalign{\smallskip}
2011		&0	&522626  &516981	&0		&   &1	&[0,1)  & 0 & 100616 & 522626 \\
2012		&1	&416101  &0	        &100616   &   &2	&[1,2) &10739   & 59673   &  416101 \\
2013		&2	&368585  &10739	&59673     &   &3	&(2,3) &9604     & 44888   &  368585 \\
2014	        &3	&329142  &9604	&44888    &  &4	&[3,4) &8475     & 37860   &  329142 \\	
2015		&4	&299830  &8475	&37860	&    &5	&[4,5) &7541     & 30658   &  299830 \\
2016 	&5	&277495  &7541	&30658    &   &6	&[5,$\infty$) &--  & 277495 & 277495 \\
\noalign{\smallskip}\hline
\end{tabular}
\end{table}

\begin{figure}[htp]\centering
\renewcommand{\figurename}{\footnotesize{Figure}}
  \includegraphics[width=0.65\textwidth]{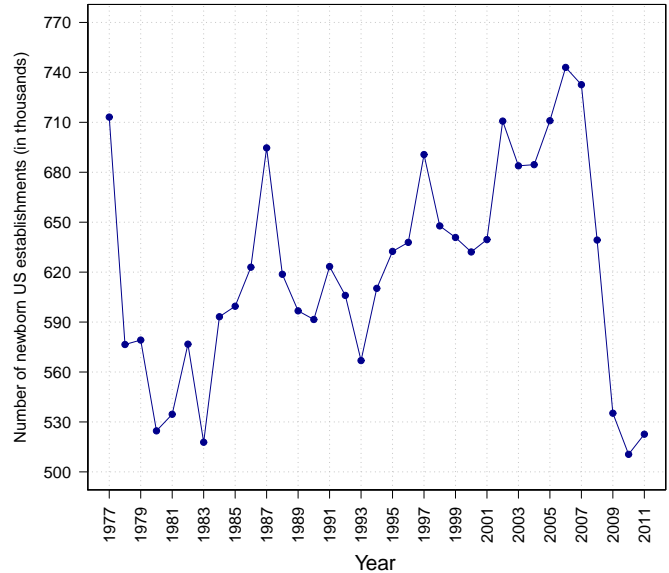}
\caption{\footnotesize Number (in thousands) of newborn US establishments, on March 12th of each year, for the period 1977-2011. Source: U.S. Census Bureau 2014, 2015 \& 2016 BDS Releases.}
\label{fig:1}
\end{figure}

In summary, our business cohort study considered the establishment as the unit of observation \cite{Audretsch1995}, considered separately 35 sequential 1-year birth cohorts (1977-2011) of US establishments, and analyzed the mortality rates for their first 5 years in business through 40 annual censuses from 1977 to 2016 published by the US Census Bureau.  Figure \ref{fig:1} shows the number (in thousands) of newborn US business (given by $Estabs$ with $Estab\_Age=0$) on March 12th of each year (in total, 21741049 establishments over the period 1977-2011 considered).

\section{Methods}
\label{sec:3}

First of all, we estimated the survival function from each of the thirty-five 1-year birth cohort considered (1977-2011). Since we had interval censored data, we obtained a non-parametric estimate for the survival function using two different methods: the Life Table method \cite{Lawless2011} and the Peto-Turnbull method \cite{Peto1973,Turnbull1976}. On the one hand, we used a modified Life Table method, taking into account the new entries of businesses with an age greater than 0 in the different cohorts. On the other hand, with respect to the Peto-Turnbull method, we used a closed-form for the non-parametric Peto-Turnbull estimator of the survival distribution, taking into account that our interval-censored business data sample is given through a set of different non-overlapping intervals, a closed-form that coincides with the well-known empirical survival function at the end-points of those intervals.

Once we obtained both non-parametric estimates of the survival function, from each of the thirty-five 1-year birth cohorts considered, we considered nine parametric probabilistic models, with support from zero to infinity, with three parameters or less, most of them well-known in the reliability analysis and actuarial literature. Then, we fitted those nine models to the $2 \times 35$ non-parametric estimates by maximum likelihood method. After that, we compared them by using the Akaike Information Criterion ($AIC$).

Finally, we analyzed the shape of the hazard function of the model/models with the best $AIC$ values, in order to test, empirically and during the first years of life, the different hypotheses proposed to explain the relationship between failure risk and business age.

\subsection{A modified Life Table method for small business mortality data}
\label{subsec:31}

The Life Table method, also called Actuarial Method \cite{Miller2011}, is a classical technique of survival analysis.
In its standard version, the time axis is divided into $k+1$ intervals $[0,a_1),\dots,[a_{j-1},a_j),\dots,[a_k,\infty)$, and the nonparametric estimator of the survival function $S(t)=\Pr(T>t)$,
is given by \cite{Lawless2011}:
\begin{equation}\label{eq:1}
\hat{P}_j=\hat{S}(a_j)=\prod_{i=1}^{j}\hat{p}_i=\prod_{i=1}^{j}(1-\hat{q}_i)=
\prod_{i=1}^{j}\left(1-\displaystyle\frac{D_i}{N_i-W_i/2}\right), \;j=1,\dots,k+1
\end{equation}
where $\hat{p}_j$, $\hat{q}_j$ are the non-parametric estimator of the conditional probability of a business of surviving beyond or of dying in the interval $I_j=[a_{j-1},a_j)$ respectively, given that the establishment survived beyond $I_{j-1}$; and $W_j\geq0$ is the number of withdrawals (active establishments reported in $N_j$ at $a_{j-1}$, lost to follow-up during $I_j$, and not reported in $N_{j+1}$ at $a_j$ as it is unknown whether they are dead or alive). It can be noted that withdrawals appear in expression (\ref{eq:1}) as $W_j/2$, assuming that withdrawals occur at the mid-point of $I_j$, on average,
following a uniform distribution \cite{Marubini2004}. We can calculate $W_j$  from the next expression:
\begin{equation}\label{eq:2}
N_{j+1}=N_j-D_j-W_j, \;j=1,\dots,k.
\end{equation}

\begin{figure}[htp]\centering
\renewcommand{\figurename}{\footnotesize{Figure}}
  \includegraphics[width=0.75\textwidth]{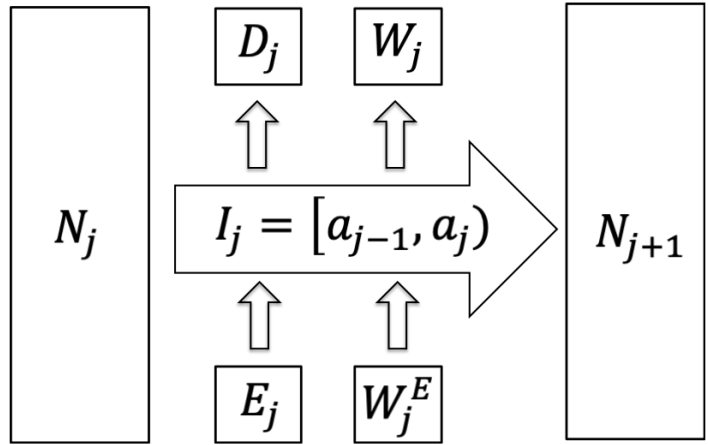}
\caption{\footnotesize Stock and Flow diagram of active establishments during a 1-year time interval $I_j=[a_{j-1},a_j)$. Stock: $N_j$ (at $a_{j-1}$) and $N_{j+1}$ at $a_j$. Outflows: $D_j$ (deaths) and $W_j$ (withdrawals). Inflows: $E_j$ (arrivals) and $W^E_j$ (remainder).}
\label{fig:2}
\end{figure}

However, in our study, BDS dataset also provides information about the number of new entrants $Estabs\_Entry$ with an age $Estab\_Age>0$ (denoted in the interval $I_j$ as $E_j$), that correspond to active US establishments not previously reported. Also, we found that there are active establishments reported in $N_{j+1}$ at $a_j$, that are not reported neither in $N_j$ at $a_{j-1}$ nor in $E_j$ during $I_j$, therefore censored establishments that did not die but might have also died during $I_j$, they are denoted by $W^E_j$. Figure \ref{fig:2} illustrates this input/output process, where $E_j$ (arrivals) and $W^E_j$ (remainder) would be equivalent on the input side to $D_j$ (deaths) and $W_j$ (withdrawals) on the output side, respectively. This input/output process can be expressed analytically as follows:
\begin{equation}\label{eq:3}
N_{j+1}=N_j+E_j+W^E_j-D_j-W_j.
\end{equation}

By taking this into consideration, we propose to include these elements (arrivals and remainder) in the expression (\ref{eq:1}) as follows:
\begin{equation}\label{eq:4}
\hat{P}_j=\prod_{i=1}^{j}\hat{p}_i=\prod_{i=1}^{j}(1-\hat{q}_i)=
\prod_{i=1}^{j}\left(1-\displaystyle\frac{D_i}{N_i+E_i-W'_i/2}\right), \;j=1,\dots,k+1
\end{equation}
where $W'_j=W_j-W^E_j$, and it has also be assumed that both $E_j$ and $W^E_j$ increase the number of establishments at risk (for half of the interval $I_j$ in both $W^E_j$ and $W_j$).
On other hand, by using Eq. (\ref{eq:3}), the expression (\ref{eq:2}) can be rewritten as 
\begin{equation}\label{eq:5}
N_{j+1}=N_j+E_j-D_j- W'_j, \;j=1,\dots,k
\end{equation}
that can be used for calculating the values of $W'_j=W_j-W^E_j$, which can be positive, negative or zero.

Following the example of the 2011 birth cohort given in Table \ref{tab:1}, Table \ref{tab:2} shows the values of $W'_j$ calculated by using the expression (\ref{eq:5}), and the nonparametric estimates of the survival function, $\hat{S}(a_j)$, obtained by the modified Life Table method given by the expression (\ref{eq:4}).

\subsection{An analytical Peto-Turnbull method for small business survival analysis}
\label{subsec:32}

The Peto-Turnbull method, an extension of the Kaplan-Meier estimator for interval-censored data \cite{Gross1992}, it is a well-known technique for non-parametric estimation of the survival distribution $S(t)=\Pr(T>t)$, based on an iterative procedure, and generally solved by using statistical software.

First, in its standard version, that procedure considers the set of intervals $[L_i,R_i],\;i=1,\dots,N$, that corresponds to the interval-censored data sample of size $N$, and that can include left-censored or right-censored as $(-\infty,R_i]$, $[L_i,\infty)$ respectively, and exact observations when $L_i=R_i$ (following the original notation in \cite{Peto1973,Turnbull1976}).

Then, from the sets of left-end points $\mathcal{L}=\{L_i,\;i=1,\dots,N\}$ and right-end points $\mathcal{R}=\{R_i,\;i=1,\dots,N\}$, the intervals $[q_j,p_j],\;j=1,\dots,m$ are derived by selecting all the possible intervals with the left-end point included in $\mathcal{L}$, with the right-end point included in $\mathcal{R}$, and that not include other values of $\mathcal{L}$ or $\mathcal{R}$ within the interval. Next, the likelihood function is expressed as follows:
\begin{equation}\label{eq:6}
L(s_1,\dots,s_m)=\prod_{i=1}^N [S(L_i)-S(R_i)]=\prod_{i=1}^N\left[\sum_{j=1}^m \alpha_{ij} s_j  \right]
\end{equation}
where $s_j=S(q_j)-S(p_j)$ and $\alpha_{ij}=\bm{1}\{[q_j,p_j]\subseteq [L_i,R_i]\}$.

Finally, the non-parametric estimate of the survival distribution is obtained from a constrained optimization problem, in which the objective function is the likelihood function $L(s_1,\dots,s_m)$ expressed in (\ref{eq:6}), that is to be maximized subject to the constraints $s_j\geq0,j=1,\dots,m$, and $\sum_{j=1}^m s_j=1$. The Peto-Turnbull non-parametric maximum likelihood estimator (NPMLE) of the survival function $\hat{S}(t)$ is given as follows:
\begin{equation}\label{eq:7}
\hat{S}(t)=
\left\{
\begin{array}{lll}
1&,\;\mbox{if}\;\;t< q_1&,\;(q_1 \not= -\infty)\\
&&\\[-3ex]
1-\displaystyle\sum_{j=1}^k \hat{s}_j&,\;\mbox{if}\;\;p_k< t< q_{k+1}&,\;k=1,\dots,m-1\\
&&\\[-2ex]
\mbox{undefined}&,\;\mbox{if}\;\;q_k\leq t \leq p_k&,\;k=1,\dots,m\\
&&\\[-2ex]
0&,\;\mbox{if}\;\;t> p_m&,\;(p_m \not= \infty).\\
\end{array}
\right.
\end{equation}

Different algorithms have been developed for solving that optimization problem and implemented
within most statistical software, as an example, the \textsf{R} package {\bf icenReg} with the modelling function \textsf{ic\_np} \cite{Gomez2009,Zhang2010,AndersonBergman2017}. However, the convergence of all those algorithms can be very slow as the sample size increases. Therefore, a closed-form for the NPMLE of the survival function $\hat{S}(t)$, at least in some common situations, would be very useful.

In this work, we used a closed-form for the NPLME of the survival function, by taking into account that our interval-censored data sample is given through a set of non-overlapping intervals, and in that case, the Turnbull's intervals are precisely those non-overlapping intervals. In our study, each of the 35 birth cohort considered is given through a set of six non-overlapping intervals $I_j=[a_{j-1},a_j),\;j=1,\dots,k+1\,\text{i.e.}\,(\{[0,1),[1,2),\dots,[5,\infty)\})$ in years (in terms of notation $m=k+1=6$), whose frequencies are given by $D_j$ (number of business deaths in a particular time interval), whose sample size is $N=\sum_{j=1}^{k+1} D_j$, and therefore,
$s_j=S(q_j)-S(p_j)=S(a_{j-1})-S(a_j),\;j=1,\dots,k+1$.
Actually, that  scenario of 'non-overlapping intervals' is not an uncommon situation. Two examples: {\it ``periodic inspections made at times} $t_1,t_2,\dots,t_m$ {\it in order to see a certain event has yet happened''} \cite{Turnbull1976}, or reading errors due to the limited resolution of the measurement instrument.

Then, under that non-overlapping intervals assumption, the likelihood function (\ref{eq:6}) can be expressed as:
\begin{equation}\label{eq:8}
L(s_1,\dots,s_{k+1})=\prod_{i=1}^N [S(L_i)-S(R_i)]=\prod_{i=1}^N\left[\sum_{j=1}^{k+1} \alpha_{ij} s_j  \right]=
\prod_{j=1}^{k+1}  s_j^{D_j}.
\end{equation}

Therefore, we can derive the non-parametric estimate of the survival distribution from a nonlinear programming problem, in which the objective function is the likelihood function $L(s_1,\dots,s_{k+1})$ expressed in (\ref{eq:8}), that is to be maximized subject to the constraints $s_j\geq0,j=1,\dots,k+1$, and $\sum_{j=1}^{k+1} s_j=1$.
The optimal solution of that problem is at
$$\left(s_1^0,\dots,s_j^0,\dots,s_{k+1}^0 \right)=
\left(D_1/N,\dots,D_j/N, \dots,D_{k+1}/N \right)$$
and the closed-form of the Peto-Turnbull non-parametric maximum likelihood estimator (NPMLE) of the survival function, $\hat{S}(t)$, under non-overlapping intervals that are a partition of the sample space, is given as follows (it can be noted that in our case, we have in expression (\ref{eq:7}) that $q_{k+1}=p_k,\;k=1,\dots,m-1$):
\begin{equation}\label{eq:9}
\hat{S}(t)=
\left\{
\begin{array}{lll}
1&,\;\mbox{if}\;\;t\leq a_0&\\
&&\\[-2ex]
1-(D_1/N+\ldots+D_j/N)&,\;\mbox{if}\;\;t= a_j&,\;j=1,\dots,k+1\\
&&\\[-2ex]
\mbox{undefined}&,\;\mbox{if}\;\;a_{j-1} < t <a_j&,\;j=1,\dots,k+1\\
\end{array}
\right.
\end{equation}
that coincides with the well-known empirical survival function at each end-point $t=a_j$ and that is undefined at all other $t\in(a_{j-1},a_j),\;j=1,\dots,k+1$.

Following the example of the 2011 birth cohort given in Table \ref{tab:1}, the next Table \ref{tab:2} shows the values of the non-parametric estimates of the survival function, $\hat{S}(a_j)$, obtained by the Peto-Turnbull method, from the closed-form proposed in expression (\ref{eq:9}).

\begin{table}[htp]\footnotesize\centering
\renewcommand{\tablename}{\footnotesize{Table}}
\setlength{\tabcolsep}{7.1pt}
\caption{\label{tab:2}\footnotesize Non-parametric estimates ($\hat{S}(a_j),j=1,\dots,6$) of the survival function ($S(t)=\Pr(T>t)$), obtained by using a modified Life Table method, and by using an analytical Peto-Turnbull method with closed-form, from the 2011 1-year birth cohort.}
\begin{tabular}{c c c c c c c c}
\hline\noalign{\smallskip}
 $j$ &  $I_j${\tiny(in years)} & $E_j$ & $D_j$ &
 $N_j$ & $W'_j$& $\hat{S}(a_j)$& $\hat{S}(a_j)$\\
    &  $[a_{j-1},a_j)$ & & & & &(Life Table)& (Peto-Turnbull)\\
\noalign{\smallskip}\hline\noalign{\smallskip}
	1	&[0,1)  & 0 & 100616 & 522626 	 &  5909 & 	0.8064& 0.8175\\
	2	&[1,2) &10739   & 59673   &  416101 &  -1418&	0.6938&	0.7092\\
	3	&[2,3) &9604     & 44888   &  368585 &   4159&	0.6110&	0.6278\\
	4	&[3,4) &8475     & 37860   &  329142 &      -73&	0.5425&	0.5591\\	
	5	&[4,5) &7541     & 30658   &  299830 &    -782&	0.4885&	0.5034\\
	6	&[5,$\infty$) &--  & 277495 & 277495 &         0&	                 0&	0\\
\noalign{\smallskip}\hline
\end{tabular}
\end{table}

\subsection{Modelling the shape of the hazard function for the newborn small businesses}
\label{subsec:33}

Once we obtained both non-parametric estimates of the survival function, $\hat{S}(a_j)$, by using a modified life table method and a closed-form of the Peto-Turnbull method, from each of the thirty-five 1-year birth cohorts considered, we fitted nine parametric probabilistic models to those estimates, in order to analyze the shape of the hazard function $h(t)$ most adequate for our data set, and then to empirically test the diverse hypotheses proposed to explain the relationship between failure risk and business age.\\

First of all, we considered nine parametric probabilistic models with support from zero to infinity, that include three or less parameters, most of them well-known in reliability analysis and actuarial science.
Table \ref{tab:3} shows, for these nine models, on the one hand, the survival function $S(x)$, and on the other hand, the hazard function $h(x)$ and their possible shapes.  In summary, we have considered:

\paragraph{A first group of four models:} the Exponential, that we denoted as EXP \cite{Coad2010};
Weibull, WEI \cite{Axtell2016};
Gamma, GAM \cite{Kleiber2003};
and Generalized Gamma, GGD \cite{Stacy1962,Glaser1980}
distributions.  All of them can exhibit a constant hazard function, allowing us to test the age-independent risk hypothesis.

The Generalized Gamma distribution, with DFR (decreasing failure rate), CFR (constant failure rate), IFR (increasing failure rate), UBT (upside-down bathtub-shape failure rate) and an additional U-shaped-bathtub behavior (BT), would correspond with all the previous hypotheses considered. In particular,  DFR would correspond with the Liability of Newness hypothesis, CFR  with the age-independent risk hypothesis, IFR  with the Liability of Senescence or the Liability Obsolescence, and UBT with the adolescence hypothesis.

In addition, the Generalized Gamma distribution includes the Gamma model
(for $\alpha=1$), the Weibull model (for $\beta/\alpha=1$), and the Exponential model
(for $\alpha=\beta=1$), as particular cases, letting us to compare those four models with three, two, two and one parameters respectively.

\paragraph{A second group of three models:}  the Lomax, also known as Pareto type II with zero location parameter, PA2 \cite{Lomax1954,Arnold2015}; Fisk, also known as log-logistic, FSK \cite{Mahmood2000};
and Burr Type XII, also known as Singh-Maddala, BUR \cite{Zimmer1998,SinghMaddala2008},
distributions.

The Burr Type XII distribution, with DFR (decreasing failure rate) and UBT (upside-down bathtub-shape failure rate), includes the Fisk model (for $\alpha=1$) and Lomax model (for $\beta=1$).

\paragraph{A third group of two models:} a member of the family of Generalized Power Law, GPL \cite{Prieto2017,Prieto2020} distributions and the Dagum, also known as inverse Burr XII, DAG \cite{Domma2002,Domma2011,Kleiber2008} distribution.

The GPL distribution, with DFR (decreasing failure rate) and UBT (upside-down bathtub-shape failure rate), includes the Lomax model (for $\beta=0$).

The Dagum distribution has DFR (decreasing failure rate), UBT (upside-down bathtub-shape failure rate), and U-shaped followed by upside-down bathtub-shape (BT+UBT) hazard rate. It can be noted, in Table \ref{tab:3}, that the shape of the hazard function, of the models considered, is easily identified from the value of their shape parameters, with the exception of the Dagum distribution \cite{Domma2002} -  in that case, we decided to plot the corresponding hazard function and compute the change-point (maximum for UBT case and minimum for BT case) with the statistical software \textsf{R}, in order to confirm its shape.\\

Secondly, we fitted those nine models to the data by the maximum likelihood method. For that, under the assumption of non-informative censoring \cite{Zhang2010,Lawless2011}, the log-likelihood function for each model is given by
\begin{equation}\label{eq:10}
\log[L(\theta)]=\sum_{j=1}^{6}D_j\log[S(a_{j-1};\bm{\theta})-S(a_j;\bm{\theta})]
\end{equation}
where $S(x)=Pr(X>x)$ is the survival function of the model considered (see Table \ref{tab:3}) with vector of parameters $\bm{\theta}$; $I_j=[a_{j-1},a_j),\;j=1,\dots,k+1$, are each of the $k+1=6$ time intervals (\{[0,1),[1,2),\dots,[5,$\infty$)\} in years); and $D_j$ is the number of business deaths during each interval $I_j$.

It can be noted, from expression (\ref{eq:9}), that $D_j=N [\hat{S}(a_{j-1})-\hat{S}(a_j)]$, where $\hat{S}(t)$ is the Peto-Turnbull non-parametric maximum likelihood estimator (NPMLE) of the survival function. Therefore, expression (\ref{eq:10}) can be rewritten as follows:
\begin{equation}\label{eq:11}
\log[L(\theta)]=\sum_{j=1}^{6}N [\hat{S}(a_{j-1})-\hat{S}(a_j)]\log[S(a_{j-1};\bm{\theta})-S(a_j;\bm{\theta})].
\end{equation}
In this paper, for comparison purposes, we explored an alternative strategy for model fitting, based in the modified Life Table method described in subsection \ref{subsec:31}, in which $\hat{S}(t)$ in expression (\ref{eq:11}) is the NPMLE of the survival function given by expression (\ref{eq:1}).\\

\begin{landscape}
\begin{table}[p]\footnotesize
\renewcommand{\tablename}{\footnotesize{Table}}
\caption{\label{tab:3}\footnotesize Survival function $S(x)$, hazard function $h(x)$ and their possible shapes, of the nine models considered: strictly decreasing (DFR), constant (CFR), strictly increasing (IFR), U-shaped-Bathtub (BT), upside-down bathtub-shaped (UBT), or U-shaped followed by upside-down bathtub-shape (BT+UBT), where
$\alpha,\beta,\sigma>0$. It can be noted that $\gamma(a,z)$ denotes the lower incomplete gamma function $\gamma(a,z)=\int_{0}^{z} t^{a-1} e^{-t} dt$.}
\begin{tabular}{llll}
\hline\noalign{\smallskip}
Model & Survival function: $S(x),\forall x>0$& Hazard function: $h(x),\forall x>0$& Possible shapes of $h(x),\forall x>0$\\
\noalign{\smallskip}\hline\noalign{\smallskip}
Exponential
& $\exp[-(x/\sigma)]$
& $1/\sigma$
& CFR\\[3ex]
Weibull
& $\exp[-(x/\sigma)^\alpha]$
& $(\alpha/\sigma)(x/\sigma)^{\alpha-1}$
& $0<\alpha<1$ (DFR), $\alpha=1$ (CFR), $\alpha>1$ (IFR)\\[3ex]
Gamma
& $1-\displaystyle\frac{\gamma[\beta,(x/\sigma)]}{\rm\Gamma(\beta)}$
& $\displaystyle\frac{(1/\sigma)(x/\sigma)^{\beta-1}\exp{[-(x/\sigma)]}}
{\rm\Gamma(\beta)-\gamma[\beta,(x/\sigma)]}$
& $0<\beta<1$ (DFR), $\beta=1$ (CFR), $\beta>1$ (IFR)\\[3ex]
& \multirow{3}{*}{$1-\displaystyle\frac{\gamma[\beta/\alpha,(x/\sigma)^\alpha]}{\rm\Gamma(\beta/\alpha)}$}
& \multirow{3}{*}{$\displaystyle\frac{(\alpha/\sigma)(x/\sigma)^{\beta-1}
\exp{[-(x/\sigma)^\alpha]}}{\rm\Gamma(\beta/\alpha)-\gamma[\beta/\alpha,(x/\sigma)^\alpha]}$}
& $\beta<1\Rightarrow$ $\alpha\leq1$ (DFR); $\alpha>1$(BT)\\
Generalized&&
& $\beta=1\Rightarrow$ $\alpha<1$ (DFR); $\alpha=1$ (CFR); $\alpha>1$(IFR)\\
Gamma&&
& $\beta>1\Rightarrow$ $\alpha<1$ (UBT); $\alpha\geq1$(IFR)\\[2ex]
\hline\noalign{\smallskip}
Lomax
& $[1+x/\sigma]^{-\alpha}$
& $\displaystyle\frac{(\alpha/\sigma)}{1+x/\sigma}$
& DFR\\[3ex]
Fisk
& $[1+(x/\sigma)^\beta]^{-1}$
& $\displaystyle\frac{(\beta/\sigma)(x/\sigma)^{\beta-1}} {1+(x/\sigma)^{\beta} }$
& $0<\beta\leq1$ (DFR), $\beta>1$ (UBT)\\[3ex]
Burr XII
& $[1+(x/\sigma)^{\beta}]^{-\alpha}$
& $\displaystyle\frac{(\alpha\beta/\sigma)(x/\sigma)^{\beta-1}} {1+(x/\sigma)^{\beta} }$
& $0<\beta\leq1$ (DFR), $\beta>1$ (UBT)\\[3ex]
\hline\noalign{\smallskip}
GPL
& $\exp\left(-\alpha\displaystyle\frac{[\log(1+x/\sigma)]^{\beta+1}}{[1+\log(1+x/\sigma)]^\beta}\right)$
& $\displaystyle\frac{(\alpha/\sigma)[\beta+1+\log(1+x/\sigma)][\log(1+x/\sigma)]^\beta}
{(1+x/\sigma)[1+\log(1+x/\sigma)]^{\beta+1}}$
& $-1<\beta\leq0$ (DFR), $\beta>0$ (UBT)\\[4ex]
Dagum
& $1-[1+(x/\sigma)^{-\beta}]^{-\alpha}$
& $\displaystyle\frac{(\alpha\beta/\sigma)(x/\sigma)^{-(\beta+1)}
[1+(x/\sigma)^{-\beta}]^{-(\alpha+1)}}
{1-[1+(x/\sigma)^{-\beta}]^{-\alpha}}$
&DFR, UBT, BT+UBT (see \cite{Domma2002})\\[2ex]
\noalign{\smallskip}\hline\noalign{\smallskip}
\end{tabular}
\end{table}
\end{landscape}

Finally, we compared those nine models by using the Akaike's information criterion, $AIC$ \cite{Akaike1974}, defined by
\begin{equation}\label{eq:12}
AIC=-2\log\ell+2K,
\end{equation}
where $\log\ell$ is the log-likelihood of the model evaluated at the maximum likelihood estimates, $K$ is the number of parameters of the model, and where a lower $AIC$ value indicates a better fit.

For that comparison, for each of the 35 birth cohorts, we obtained the minimum value of $AIC$, denoted as
$AIC_{min}$, and then, we computed for each of the 9 models its value of $\Delta=AIC-AIC_{min}$ \cite{Burnham2004,Burnham2011}. After that, we classified those 9 models in three group, for each birth cohort: models with the best fit, given by $\Delta\leq 2$; models with relative little support, given by $2<\Delta\leq 20$; and models with no empirical support, given by $\Delta>20$. We obtained that classification for each of the two techniques of survival analysis used: Lifetable and Peto-Turnbull methods.  As far as we know, there is no an agreement with respect to the previous intervals of $\Delta$, so we considered the first interval $\Delta\leq 2$ as a way, not only for finding the best models but also for comparing models with one parameter of difference  in which $AIC$ penalizes the model with one more parameter with $2K=2$, and we considered the third interval $\Delta>20$ as the worst case of the two options proposed in
\cite{Burnham2004} and \cite{Burnham2011}. A direction for future research could be based on analyzing the optimum cutoff rules for that comparison.

\section{Empirical results}
\label{sec:4}

\subsection{Non-parametric estimation of the survival rates of US newborn establishments}

What are the survival rates of US business during their five years of life?\\

\begin{figure}[htp]\centering
\renewcommand{\figurename}{\footnotesize{Figure}}
  \includegraphics[width=0.7\textwidth]{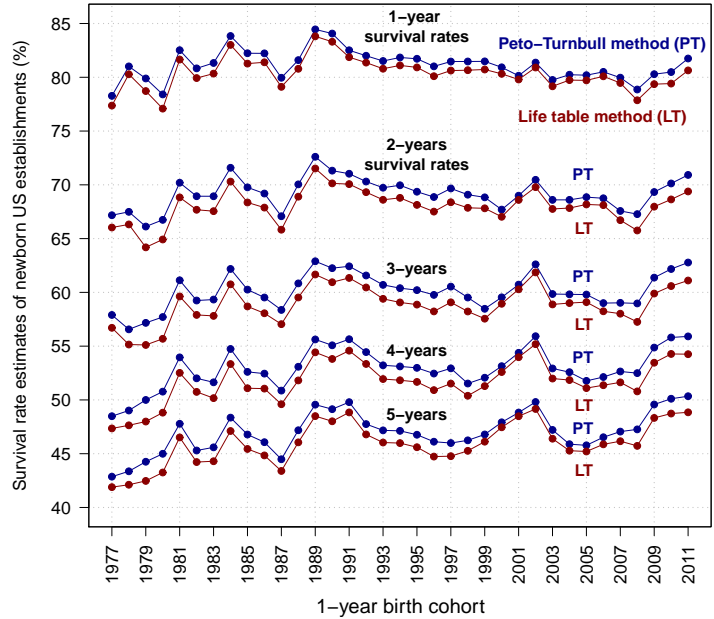}
\caption{\footnotesize Survival rate estimates (in \%) of newborn establishments in US, from 1-year to 5-years survival, for each 1-year birth cohort from period 1977-2016. Source: U.S. Census Bureau 2014, 2015 \& 2016 BDS Releases.}
\label{fig:3}
\end{figure}

Figure \ref{fig:3} shows the survival rate estimates (in \%) of newborn establishments in US, from 1-year to 5-years survival, for each 1-year birth cohorts from the period 1977-2016, obtained by using a
modified Life Table method (by taking into account the new entries of businesses with an age greater than 0 in the different cohorts) and also by using a closed-form for the non-parametric Peto-Turnbull estimator of the survival distribution (by also considering that our interval-censored business data sample is given through a set of different non-overlapping intervals).

It can be noted that, depending on the birth year and on the estimation method (life table method, LT, shows lower values), 
approximately 77\%-84\% of newborn establishments survived 1 year or more,
64\%-73\% survived 2 years o more, 55\%-63\% survived 3 years of more, 
47\%-56\% survived 4 years or more, and only 42\%-50\% survived the first five years of life.

\subsection{Parametric estimation and model selection in the analysis of new establishments survival}

Which, of the nine parametric models considered, does seem to fit better the data?\\

\begin{figure}[htp]\centering
\renewcommand{\figurename}{\footnotesize{Figure}}
  \includegraphics[width=1\textwidth]{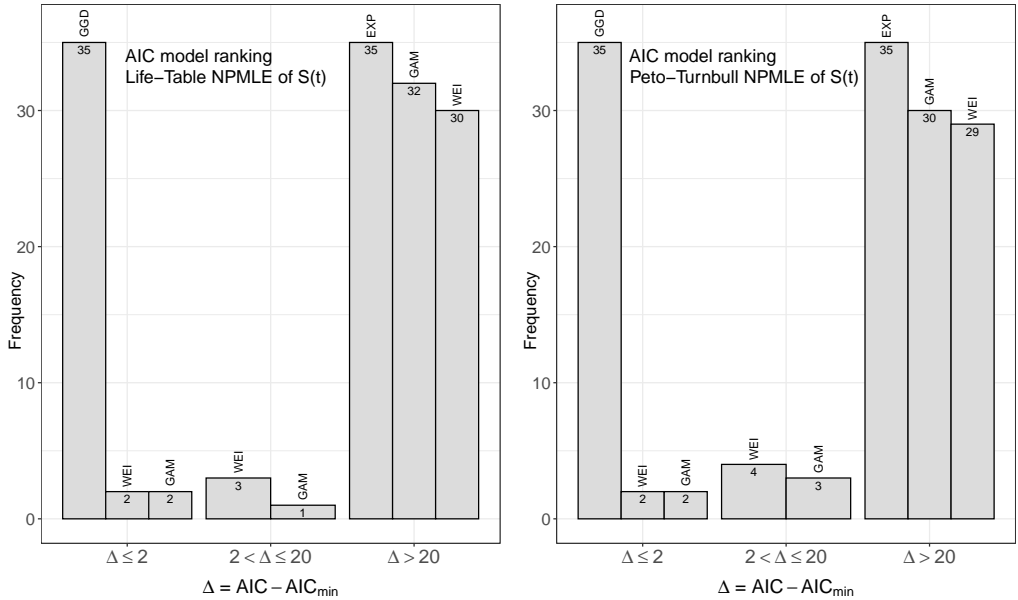}
\caption{\footnotesize AIC model ranking. Number of times that each of the models considered fits the data best ($\Delta=AIC-AIC_{min}<=2$), that has relatively little support ($2<\Delta<=20$), or that has no empirical support ($\Delta>20$), in the period 1977-2016 (35 one-year birth cohorts in total). Models considered: Exponential (EXP), Weibull (WEI), Gamma (GAM) and Generalized Gamma (GGD) distributions.}
\label{fig:4}
\end{figure}

Figure \ref{fig:4} corresponds with the first group of four nested models:
the Exponential (EXP), Weibull (WEI), Gamma (GAM) and Generalized Gamma (GGD) distributions. It shows the number of times (frequency) that each of those four models fits the data best ($\Delta=AIC-AIC_{min}<=2$), that has relatively little support ($2<\Delta<=20$), or that has no empirical support ($\Delta>20$), in the period 1977-2016 (35 birth cohorts in total). That AIC model ranking, on the one hand, is obtained by using a modified Life Table method (on the left panel), and on the other hand, by using a closed-form for the non-parametric Peto-Turnbull estimator of the survival distribution (on the right panel). It can be noted that the Generalized Gamma (GGD) distribution fits the data best in all the 35 one-year birth cohorts and with both methods, and that the Exponential, Weibull and Gamma distributions have no empirical support in 35, 32, 30 birth cohorts respectively (by using the Life Table method), and in 35, 30, 29 birth cohorts respectively (by using the Peto-Turnbull method) - in fact,  only in 2 of the 35 birth cohorts, the Weibull and Gamma distributions tie with the GGD model for first place.

\begin{figure}[htp]\centering
\renewcommand{\figurename}{\footnotesize{Figure}}
  \includegraphics[width=1\textwidth]{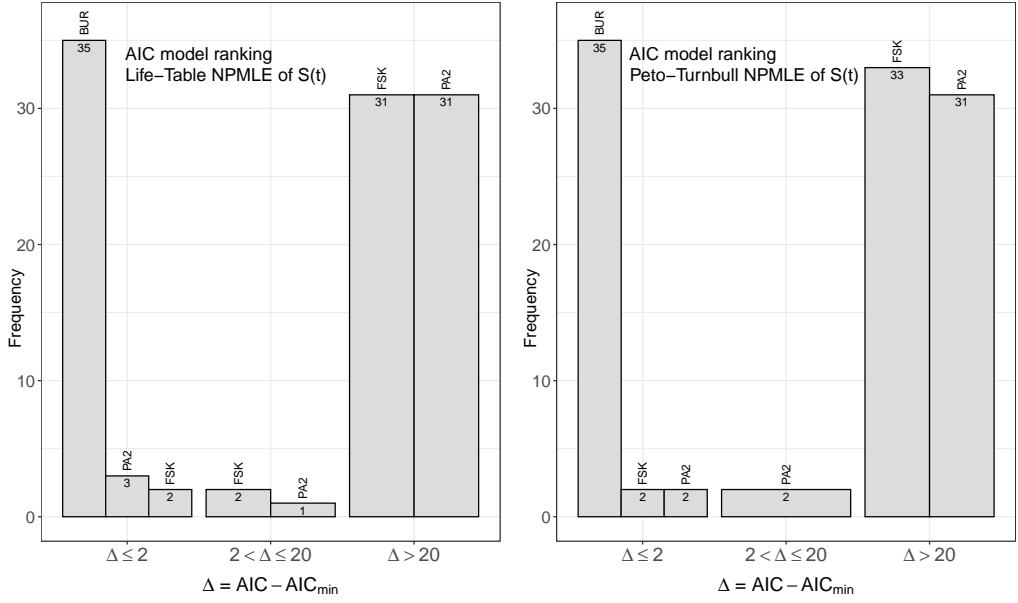}
\caption{\footnotesize AIC model ranking: Lomax (PA2), Fisk (FSK) and Burr type XII (BUR) models.}
\label{fig:5}
\end{figure}

Figures \ref{fig:5} corresponds with the following group of three nested models: the Lomax (PA2), Fisk (FSK) and Burr type XII (BUR) distributions. It shows the AIC model ranking obtained by the Life Table method (on the left panel) and by the Peto-Turnbull method (on the right panel) from those three models. It can be observed that the Burr type XII (BUR) distribution fits the data best in all the 35 one-year birth cohorts and with both methods, and that the Lomax and Fisk distributions have no empirical support in 31 birth cohorts (by using the Life Table method), and in 33, 31 birth cohorts respectively (by using the Peto-Turnbull method).

\begin{figure}[htp]\centering
\renewcommand{\figurename}{\footnotesize{Figure}}
  \includegraphics[width=1\textwidth]{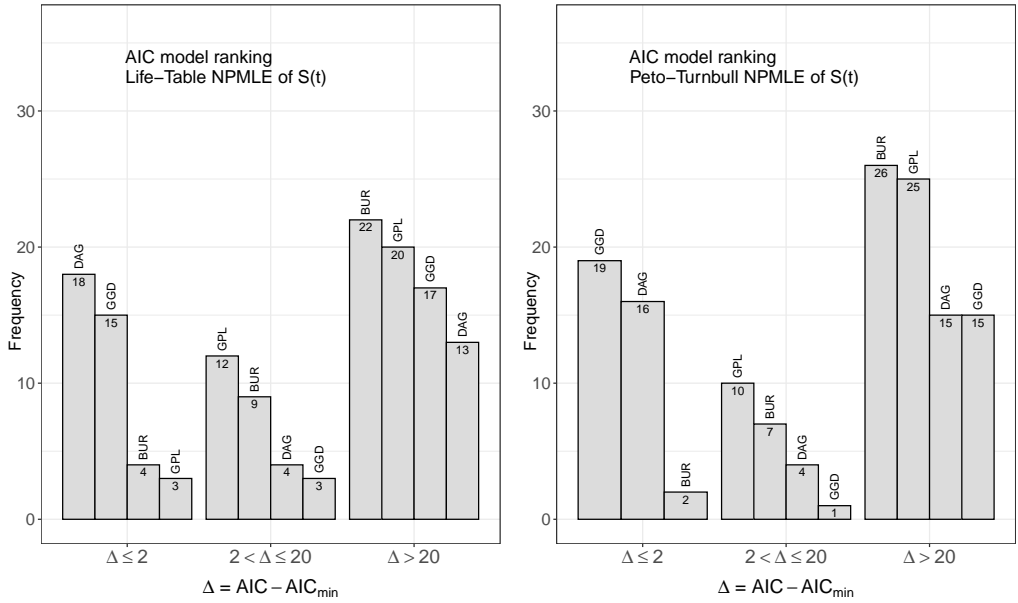}
\caption{\footnotesize AIC model ranking: Generalized Gamma (GGD), Burr type XII (BUR), Generalized Power Law (GPL) and Dagum (DAG) distributions.}
\label{fig:6}
\end{figure}

Figures \ref{fig:6} corresponds with the group of non-nested 3-parameters models considered, that is composed of the following four models: the Generalized Gamma (GGD), the Burr type XII (BUR), a Generalized Power Law (GPL) and the Dagum (DAG) distributions. In combination with Table \ref{tab:4},
it can be noted that no model seems to fit the data best in all the 35 one-year birth cohorts considered: \\
{\it The GGD model} fits the data best in 15 birth cohorts (14 of them alone) in lifetable case (on the left), and in 19 cohorts (18 of them alone) in Peto-Turnbull case (on the right panel).\\
{\it The DAG model} fits the data best in 18 cohorts (15 of them alone) in lifetable case (on the left panel),
and in 16 cohorts (14 of them alone) in Peto-Turnbull case (on the right panel).\\
{\it The BUR model} fits the data best in 4 cohorts (2 of them alone) in lifetable case (on the left panel),
and in 2 cohorts (1 of them alone) in Peto-Turnbull case (on the right panel). That model has relative little support ($2<\Delta\leq 20$) in 9 cohorts (in lifetable case) and in 7 cohorts (in Peto-Turnbull case).\\
{\it The GPL model} fits the data best in 3 cohorts (1 of them alone) in lifetable case (on the left panel) and has relative little support in 12 cohorts (lifetable) and in 10 cohorts (Peto-Turnbull).\\
Although the Generalized Gamma and Dagum distributions, both together, fit the data best in almost all the cohorts, BUR and GPL models fit best in a few cohorts and cannot be dismissed in some of the other groups.

In summary, none of the nine parametric models considered seems to fit better the data than the rest of the models. However the Generalized Gamma and Dagum models, both together, seems to be the best models, and the analysis of new establishments survival during the first years of life can be better described based on both models together with the Burr type XII and Generalized Power Law models.  A line for future research could be based on the search for a model that fit better all the 35 birth cohorts, by following the principle of parsimony.

\subsection{Empirical analysis of the diverse hypotheses proposed for explaining the relationship between failure risk and business age, in the first years of life}

How does small business die during their first years of life? What shape does the hazard function have in the first five years in market?\\

Table \ref{tab:4} shows, for each birth cohort (from 1977 to 2011), and for each technique of survival analysis used (Lifetable on the left and Peto-Turnbull on the right): the models that fit the data best ($\Delta=AIC-AIC_{min}<=2$) and the models that has relatively little support ($2<\Delta<=20$), ranked by the value of $\Delta$; the value of $\Delta$; the shape of the hazard function $h(x)$ according to the values of the parameter estimates obtained by maximum likelihood (see table \ref{tab:3}), and finally, in the case of BT or UBT shapes, the value $x_0$ of the change-point expressed in months (the value of the maximal or the minimal point respectively).

\begin{figure}[htp]\centering
\renewcommand{\figurename}{\footnotesize{Figure}}
  \includegraphics[width=1.0\textwidth]{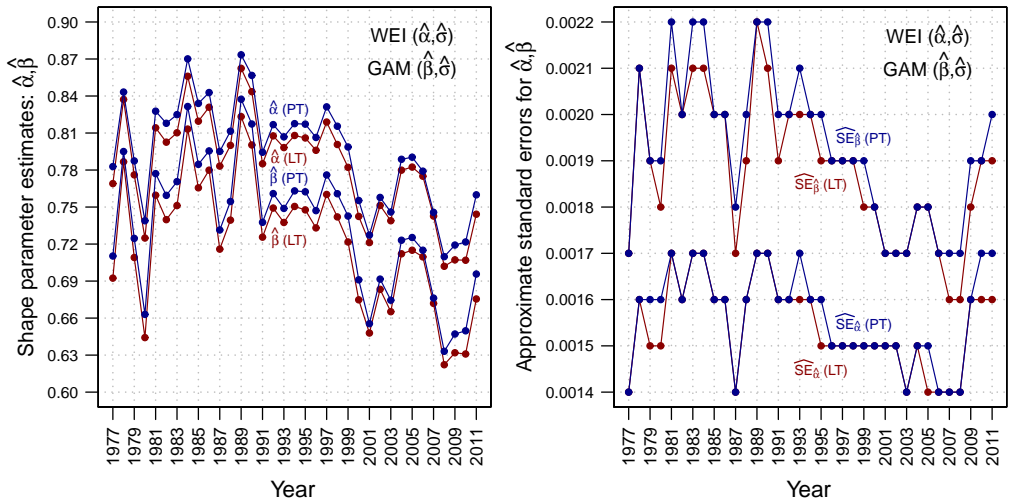}
\caption{\footnotesize Shape parameter estimates for the Weibull and Gamma distributions, fitted by maximum likelihood estimation for each technique of survival analysis considered (LT: Lifetable method, PT: Peto-Turnbull method), and corresponding standard errors, in each of the thirty five 1-year birth cohorts from period 1977-2016.}
\label{fig:7}
\end{figure}

First of all, it can be noticed that the age-independent risk hypothesis has no empirical support in the first five years of life of a business. That hypothesis would be favored by the exponential model (but it was the worst model in our comparison, with the worst values of $AIC$ and no empirical support in any cohort), by the Weibull model with $\alpha=1$, by the Gamma with $\beta=1$, or by the Generalized Gamma model  with $\alpha=\beta=1$. Figure \ref{fig:7} shows the values of $\alpha$ and $\beta$ estimates for WEI and GAM models, and their standard errors.

Secondly, it can be noted that the Liability of Senescence and the Liability of Obsolescence hypotheses have no empirical support in the first five year in market of a business. We did not find out, during those first five years, a strictly increasing (IFR) shape in any birth cohort. That IFR shape could be modeled by the Weibull distribution with $\alpha>1$ or by the Gamma distribution with $\beta>1$ (or with the corresponding generalization by the GGD model), and it can be noted in figure \ref{fig:7} that both shape parameter estimates were less than one in each cohort. 

Thirdly, we did not find out a U-shaped followed by upside-down bathtub-shape (BT+UBT) shape in any birth cohort. 

Fourthly, we found a U-shaped-bathtub behavior (BT) shape in two years, both modeled by the GGD model with shape parameter estimates and their standard errors as follows:
($\alpha,\beta$)=(1.18,0.74) and ($\hat{SE}_{\hat{\alpha}}$,$\hat{SE}_{\hat{\beta}}$)=(0.032,0.005) in 1997;
($\alpha,\beta$)=(1.20,0.69) and ($\hat{SE}_{\hat{\alpha}}$,$\hat{SE}_{\hat{\beta}}$)=(0.035,0.005) in 2004.
However, in both cases, the change-point $x_0$ (minimal point) went beyond the first five years of life analyzed in this paper (109.3 and 129.6 months respectively), and in addition, the BT shape was only observed with the Peto-Turnbull method and not with the Life Table Method (in which we have a DFR shape with the same GGD model). Thus, a line for future research could be based on the validation that BT shape beyond the first years of life of a business via higher resolution data of those years.

Fifthly, we found an upside-down bathtub-shaped (UBT or unimodal) behavior in 13 of the 35 birth cohorts, 8 of them by using both techniques (Lifetable and Peto-Turnbull methods) and with all the models with the best fit ($\Delta\leq 2$) or with relative little support ($2<\Delta<=20$). The change-point $x_0$ (maximal point in this case) was reached during the first six months in all those birth cohorts. Therefore, the Liability of Adolescence hypothesis, with an early change point in the first months, seems to be empirically supported in those years.
As we used data with 1-year resolution in the range of age of 0-5 years, a direction for future research could be based on the computation of this change-point by using higher resolution data.

Finally, we found that DFR shape is the most common behavior in the hazard function during the first five year of a business in market. Table \ref{tab:4} shows that the risk of business failure declines with the firm age during those first five years, with the exception of the first months of some years in which the risk can rise. Thus, the results obtained favor the Liability of Newness hypothesis in those five first years, complemented with the Liability of Adolescence with a change-point in the first months in some years. Therefore, aging confers an advantage for living longer due to a decreasing influence with time of some negative factors affecting businesses. 

\afterpage{
\clearpage
\begin{table}[htp]\footnotesize
\renewcommand{\tablename}{\footnotesize{Table}}
\setlength{\tabcolsep}{7pt}
\caption{\label{tab:4}\footnotesize Shape of the hazard function $h(x)$, and $\Delta=AIC-AIC$ value, of the best models ($\Delta\leq 2$) and other models with relative little support ($2<\Delta<=20$), by using Life table and Peto-Turnbull methods, of each of the birth cohort considered (1977-2011). Change-point $x_0$, expressed in months, for the case of UBT and BT shapes (maximal or minimal point respectively).}
\begin{tabular}{c c c c c c c c c c c c}
\hline\noalign{\smallskip}
&&\multicolumn{4}{c}{Life Table Method}&& \multicolumn{4}{c}{Peto-Turbull Method}\\
\cline{3-6}\cline{8-11}\noalign{\smallskip}
Year && Best Model & $\Delta$ & Shape& $x_0$&&
	   Best Model & $\Delta$ & Shape& $x_0$\\
\noalign{\smallskip}\hline\noalign{\smallskip}
1977&&	DAG&	0	  &DFR	&		&&DAG	&0		&DFR	&\\
\noalign{\smallskip}\hline\noalign{\smallskip}
\multirow{3}{*}{1978}&&BUR&0 &UBT&	5.0	&&BUR	&0		&UBT	&4.2\\
	&&	DAG&	18.3 &UBT	&	4.7	&&DAG	&1.3		&UBT	&4.3\\
 	&&	 	&	 	  & 		& 		&&GPL	&17.7	&UBT	&4.2\\
\noalign{\smallskip}\hline\noalign{\smallskip}
\multirow{3}{*}{1979}&&DAG&0	 &UBT&	3.6	&&DAG	&0		&UBT	&1.1\\
	&&	GPL&	18.3  &UBT	&	3.2	&&GPL	&3.6		&DFR	&\\
	&&		&		   &		&		&&BUR	&4.6		&DFR	&\\
\noalign{\smallskip}\hline\noalign{\smallskip}
\multirow{2}{*}{1980}&&DAG&0	  &DFR&		&&DAG	&0		&DFR	&\\
	&&	GPL&	15.2  &DFR	&		&&GGD	&14.0	&DFR	&\\
\noalign{\smallskip}\hline\noalign{\smallskip}
1981&&	GGD&	0	  &UBT	&	2.3	&&GGD	&0		&UBT	&1.3\\
\noalign{\smallskip}\hline\noalign{\smallskip}
1982&&	GGD&	0	  &DFR	&		&&GGD	&0		&DFR	&\\
\noalign{\smallskip}\hline\noalign{\smallskip}
\multirow{3}{*}{1983}&&DAG&0&DFR&		&&DAG	&0		&DFR	&\\
	&&	GPL&	12.3  &DFR	&		&&GPL	&11.6	&DFR	&\\
	&&	BUR&	15.9  &DFR	&		&&BUR	&17.1	&DFR	&\\
\noalign{\smallskip}\hline\noalign{\smallskip}
1984&&	GGD&	0	  &UBT	&	3.3	&&GGD	&0		&UBT	&2.6\\
\noalign{\smallskip}\hline\noalign{\smallskip}
\multirow{3}{*}{1985}&&DAG&0&DFR&		&&DAG	&0		&DFR	&\\
	&&	GPL&	7.1	  &DFR	&		&&GPL	&8.2		&DFR	&\\
	&&	BUR&	7.9	  &DFR	&		&&BUR	&9.7		&DFR	&\\
\noalign{\smallskip}\hline\noalign{\smallskip}
\multirow{4}{*}{1986}&&GPL&0&UBT&	2.1	&&GGD	&0		&UBT	&3.0\\
	&&	BUR&	3.5	  &UBT	&	1.7	&&		&		&		&\\
	&&	GGD&	3.8   	  &UBT	&	2.7	&&		&		&		&\\
	&&	DAG&	8.0	  &UBT	&	3.6	&&		&		&		&\\
\noalign{\smallskip}\hline\noalign{\smallskip}
1987&&	GGD&	0	  &DFR	&		&&GGD	&0		&DFR	&\\
\noalign{\smallskip}\hline\noalign{\smallskip}
1988&&	DAG&	0	  &DFR	&		&&DAG	&0		&DFR	&\\
\noalign{\smallskip}\hline\noalign{\smallskip}
\multirow{3}{*}{1989}&&DAG&0&UBT&	5.7	&&DAG	&0		&DFR	&\\
	&&	BUR&	1.9	   &UBT	&	3.6	&&BUR	&7.0		&UBT	&2.1\\
	&&	GPL&	5.1	  &UBT	&	3.7	&&GPL	&8.1		&UBT	&2.2\\
\noalign{\smallskip}\hline\noalign{\smallskip}
\multirow{3}{*}{1990}&&DAG&0&UBT&	5.7	&&DAG	&0		&UBT	&6.5\\
	&&	GPL&	8.00	  &UBT	&	5.7	&&GPL	&2.6		&UBT	&5.3\\
	&&		&		  &		&		&&BUR	&15.0	&UBT	&5.0\\
\noalign{\smallskip}\hline\noalign{\smallskip}
\multirow{2}{*}{1991}&&GGD&0&UBT&0.6		&&GGD	&0		&DFR	&\\
	&&	BUR&	18.4  &DFR	&		&&BUR	&17.9	&DFR	&\\
\noalign{\smallskip}\hline\noalign{\smallskip}
1992&&	GGD&	0	  &DFR	&		&&GGD	&0		&DFR	&\\
\noalign{\smallskip}\hline\noalign{\smallskip}
\multicolumn{11}{r}{Continued on next page}\\
\hline\noalign{\smallskip}
\end{tabular}
\end{table}
\thispagestyle{empty} 
\clearpage
}
\afterpage{
\clearpage
\begin{table}[p]\footnotesize
\renewcommand{\tablename}{\footnotesize{Table 4 (cont.)}}
\setlength{\tabcolsep}{6.7pt}
\caption*{\label{tab:4cont}}
\begin{tabular}{c c c c c c c c c c c c}
\hline\noalign{\smallskip}
&&\multicolumn{4}{c}{Life Table Method}&& \multicolumn{4}{c}{Peto-Turbull Method}\\
\cline{3-6}\cline{8-11}\noalign{\smallskip}
Year && Best Model & $\Delta$ & Shape&	$x_0$&&
	     Best Model & $\Delta$ & Shape& 	$x_0$\\
\noalign{\smallskip}\hline\noalign{\smallskip}
\multirow{4}{*}{1993}&&DAG&0&DFR&		&&DAG	&0		&DFR	&\\
	&&	GPL&	2.9	 &DFR	&		&&GGD	&0.2		&DFR	&\\
	&&	GGD&	10.0  &DFR	&		&&		&		&		&\\
	&&	BUR&	15.4  &DFR	&		&&		&		&		&\\
\noalign{\smallskip}\hline\noalign{\smallskip}
\multirow{3}{*}{1994}&&DAG&0&DFR&		&&DAG	&0		&DFR	&\\
	&&	GPL&	14.7   &DFR	&		&&GPL	&18.1	&DFR	&\\
	&&	BUR&	19.5	  &DFR	&		&&		&		&		&\\
\noalign{\smallskip}\hline\noalign{\smallskip}
1995&&	GGD&	0	  &UBT	&1.1		&&GGD	&0		&UBT	&0.2\\
\noalign{\smallskip}\hline\noalign{\smallskip}
1996&&	GGD&	0	  &DFR	&		&&GGD	&0		&DFR	&\\
\noalign{\smallskip}\hline\noalign{\smallskip}
1997&&	GGD&	0	  &DFR	&		&&GGD	&0		&BT		&109.3\\
\noalign{\smallskip}\hline\noalign{\smallskip}
\multirow{3}{*}{1998}&&DAG&0&DFR&		&&DAG	&0		&DFR	&\\
	&&	GPL&	9.9	  &DFR	&		&&		&		&		&\\
	&&	BUR&	11.9   &DFR	&		&&		&		&		&\\
\noalign{\smallskip}\hline\noalign{\smallskip}
1999&&	BUR&	0	  &UBT	&3.6		&&BUR	&0		&UBT	&3.7\\
\noalign{\smallskip}\hline\noalign{\smallskip}
2000&&	DAG&	0	  &UBT	&3.7		&&DAG	&0		&UBT	&4.0\\
\noalign{\smallskip}\hline\noalign{\smallskip}
\multirow{2}{*}{2001}&&GGD&0&DFR&		&&GGD	&0		&DFR	&\\
	&&	GPL&	11.9	  &DFR	&		&&GPL	&17.4	&DFR	&\\
\noalign{\smallskip}\hline\noalign{\smallskip}
2002&&	GGD	&	0	  &DFR	&		&&GGD	&0		&DFR	&\\
\noalign{\smallskip}\hline\noalign{\smallskip}
2003&&	DAG&	0	  &DFR	&		&&DAG	&0		&DFR	&\\
\noalign{\smallskip}\hline\noalign{\smallskip}
2004&&	GGD&	0	  &DFR	&		&&GGD	&0		&BT		&129.6\\
\noalign{\smallskip}\hline\noalign{\smallskip}
2005&&	DAG&	0	  &DFR	&		&&DAG	&0		&DFR	&\\
\noalign{\smallskip}\hline\noalign{\smallskip}
2006&&	DAG&	0	  &DFR	&		&&DAG	&0		&DFR	&\\
\noalign{\smallskip}\hline\noalign{\smallskip}
\multirow{3}{*}{2007}&&DAG&0&DFR&		&&GGD	&0		&UBT	&0.4\\
	&&	BUR&	2.4 	  &DFR	&		&&		&		&		&\\
	&&	GPL&	11.5   &DFR	&		&&		&		&		&\\
\noalign{\smallskip}\hline\noalign{\smallskip}
\multirow{4}{*}{2008}&&DAG&0&DFR&		&&GGD	&0		&DFR	&\\
	&&	GPL&	0.1	  &DFR	&		&&BUR	&3.1		&DFR	&\\
	&&	BUR&	0.1	  &DFR	&		&&GPL	&5.4		&DFR	&\\
	&&	GGD&	19.0  &UBT	&	0.10	&&DAG	&7.4		&DFR	&\\
\noalign{\smallskip}\hline\noalign{\smallskip}
\multirow{4}{*}{2009}&&GGD&0&DFR&		&&GGD	&0		&DFR	&\\
	&&	BUR&	4.1	  &DFR	&		&&GPL	&6.8		&DFR	&\\
	&&	GPL&	7.6	  &DFR	&		&&DAG	&10.9	&DFR	&\\
	&&	DAG&	8.9	  &DFR	&		&&		&		&		&\\
\noalign{\smallskip}\hline\noalign{\smallskip}
\multirow{2}{*}{2010}&&GGD&0&DFR&		&&GGD	&0		&DFR	&\\
	&&	DAG&	17.4	  &DFR	&		&&DAG	&14.2	&DFR	&\\
\noalign{\smallskip}\hline\noalign{\smallskip}
\multirow{3}{*}{2011}&&GPL&0&DFR &		&&GGD	&0		&DFR	&\\
	&&	DAG&	0.4	  &DFR	&		&&DAG	&3.3		&DFR	&\\
	&&	GGD	&	1.5	  &DFR	&		&&		&		&		&\\
\hline\noalign{\smallskip}
\end{tabular}
\end{table}
\thispagestyle{empty} 
\clearpage
}

\section{Conclusions}
\label{sec:5}

In this paper, we analyzed the relationship between failure risk (risk of death) and firm age, during the first five year in market.

For that reason, we used US newborn establishment data from the United States Census Bureau's Business Dynamics Statistics (BDS) database, with 1-year resolution in the range of age of 0-5 years, in the period 1977-2016 - in total, 35 birth cohorts were analyzed separately.

We explored the adaptation of classical techniques of survival analysis (the Life Table and Peto-Turnbull methods) to the business survival analysis. We used a modified Life Table method, by taking into account the new entries of businesses with an age greater than 0 in the different cohorts. In addition, we used a closed-form for the non-parametric Peto-Turnbull estimator of the survival distribution, by considering that our interval-censored business data sample is given through a set of different non-overlapping intervals. To the best of our knowledge, the approach introduced in this work has not been previously considered in the analysis of the risk of death in newborn businesses.

In our analysis, we considered nine parametric probabilistic models, most of them well-known in reliability analysis and actuarial science. These models include different shapes in the hazard function. Their parameters were estimated by the method of maximum likelihood method and they were compared in terms of the Akaike's information criterion. Finally, the shape of the models with the better goodness of fit measures were analyzed. The followings findings were observed:

We found that, in the period 1977-2016, from approximately  77\% to 84\% of newborn establishments survived 1 year or more, from 64\% to 73\% survived 2 years o more, from 55\% to 63\% survived 3 years of more, from 47\% to 56\% survived 4 years or more, and only from 42\% to 50\% survived the first five years in market.

We found that {\it the liability of senescence} and {\it the liability of obsolescence} hypotheses, both suggesting that the risk of business failure rises with the firm age, have no empirical support for the first five years of life of a business, in all the period 1977-2016 considered.

We found that the age-independent risk hypothesis, that suggests that the risk of business failure is approximately constant, independent of the business age, has no empirical support for the first five years of life of a business either, in all the period 1977-2016 considered.

We found empirical evidence to support the {\it The Liability of Newness} hypothesis that suggests that the risk of business failure declines with the firm age. In addition, we have found that {\it the liability of adolescence} has empirical support in some of the years considered, with an early change-point (maximal hazard rate) in the first six months of life.

In summary, we found that newborn businesses seem to have a decreasing failure rate with the age during the first five years in market, with the exception of the first months of some years in which the risk can rise.

\section*{Acknowledgements}
Faustino Prieto acknowledges funding by the Jos\'e Castillejo Program (Grant number CAS17/00461, Ministerio de Educaci\'on, Cultura y Deporte, Programa Estatal de Promoci\'on de Talento y su Empleabilidad en I+D+i, Subprograma Estatal de Movilidad, del Plan Estatal de Investigaci\'on Cient\'{\i}fica y
T\'ecnica y de Innovaci\'on 2013-2016). 
Faustino Prieto also acknowledges the Faculty of Business and Economics and the Centre for Actuarial Studies at the University of Melbourne for their special support.
Research partially carried out while Calderin-Ojeda visited University of Cantabria as part of his Special Study Program leave (University of Melbourne).
Jos\'e Mar\'{\i}a Sarabia and Faustino Prieto thanks to Ministerio de Ciencia e Innovaci\'on, project PID2019-105986GB-C22, for partial support of this work.

\bibliographystyle{plain}

\end{document}